\documentclass[10pt]{revtex4}
\usepackage{amsmath,graphicx}
\begin{document}

\title{Kepler versus Akaike}
\author{Tomasz Stachowiak}
\affiliation{Astronomical Observatory, Jagiellonian University\\
ul. Orla 171, 30-244 Krak{\'ow}, Poland}
\email{toms@oa.uj.edu.pl}

\begin{abstract}
I use the example of the Earth's orbit to illustrate the principle behind the
Akaike Information Criterion, and refute the misconception that the
criterion, by definition, discards more complex models in favour of simpler ones.
\end{abstract}

\maketitle

\section{Introduction}

AIC \cite{Akaike} is a model selection criterion, which takes into account how
well a model explains the data, but also if the model is not too
complicated.
This is intuitively understood just by looking at the formula
\begin{equation}
    \rm{AIC} = 2k - 2\ln L,\label{aka}
\end{equation}
where $k$ is the number of parameters in the fitted model, and $L$ is the
maximum value of the likelihood function. Thus, the better the fit, the lower
the value of AIC. On the other hand, the number of the model's parameters is
considered to directly signify its complexity, increasing the value of AIC. The
absolute value of AIC is not used, as it is the difference for a pair of models
that matters:
$\rm{AIC}_1 - \rm{AIC}_2 > 0$ indicates that the second model is better than
the first. The detailed mathematics of the criterion is not required here;
a more complete introduction can be found for example in \cite{Burnham}.

One may ask, if such a dependence on $k$ is not biased somehow -- to promote
models with as few parameters as possible or, perhaps, perfect agreement with
the data points. This highly subjective question remains open, as there are
many different information criteria on the market, and the concepts of
simplicity of elegance of models have not to date been unanimously defined
\cite{Dowe}.

Nevertheless, it is possible to conduct an anachronic experiment -- to test the
test itself -- by applying it to a solved problem in which a new, more
complicated theory undoubtedly replaced the older one. In other words: if the
past scientists had used model selection criteria, would physics have stopped
at the stage of naive yet elegant theories, only to achieve simplicity
instead of agreement with experiment?

The idea of applying AIC to a classical problem (the shape of the Earth's
orbit in this case) was born when a referee of \cite{Szydlowski} gave what he
thought was a {\it counterexample} to applicability of such a criterion. It
turns out that the calculations show exactly the opposite.

\section{The ``experiment''}

The difference between an ellipse and a circle is a clear example of the
complexity-accuracy trade-off. Is it really necessary to include two new
parameters: the eccentricity and the anomaly of the perihelion? Why not stick to
a simple circular orbit which is roughly the same?

Imagine we measure the distance to the Sun daily. If one is interested in its
relative value it suffices to measure the position $\varphi$ of the Sun, which
translates into angular velocity $\varphi'(t)$. Assuming next that the
field velocity is constant, one readily obtains the relative distance
\[\frac{r_i}{r_j}= \sqrt{\frac{\varphi'(t_j)}{\varphi'(t_i)}}.\]

Of course, this is partially a thought experiment, so that we have to ignore
some practical questions like how exactly the angles are to be measured. On the
other hand, we obviously assume radar has not yet been invented and only allow
for astrometric means.

Say we perform 50 observations a day for the whole year, averaging so that there
are 365 data points consisting of pairs $(\varphi_i,r_i)$. If the orbit is
elliptic then the change of $\varphi$ will vary with each day. To remedy this
one could say that the observing takes places at different times and since the
angle is also measured the data is accordingly reduced. Also, the differences
will not be big for our orbit, and it is necessary to estimate the total errors
arbitrarily anyway -- this is a thought experiment after all. Accordingly, the
anomaly $\varphi$ will be taken as known exactly and distributed evenly between
$0$ and $2\pi$.

The above setup allows us to use the simple orbit equation for the 
radius
\begin{equation}
    r = \frac{p}{1+e\cos(\varphi-\varphi_0)},
\end{equation}
where $e$ is the eccentricity, $\phi_0$, is the perihelion anomaly, and $p$ is
the distance at $\varphi=\varphi_0+\pi/2$. For a circle $e=0$, so there are
$k_1=0$ parameters. The other orbit requires $k_2=2$ parameters. How does one
choose the day to use the corresponding radius as reference, and get rid of
$p$? Since $\varphi_0$ is unknown, we could for example take the mean
reciprocal radius
\[ \langle r^{-1}\rangle = p^{-1}\langle 1+e\cos(\varphi-\varphi_0)\rangle
=p^{-1}. \]
The left hand side is obtained from the observations, and the right hand side
is an integral over $d\varphi$ -- a justified approximation taking into account
the number of the data points, and the hypothetical nature of the experiment.

Obviously, the data has to be simulated, to be as expected for an elliptic orbit
with Gaussian errors $\epsilon(0,\sigma)$ of mean $0$ and standard deviation of
$\sigma=0.5$ for a single observation. Note that 
the distance is relative, so the error corresponds to uncertainty of half of
the orbits (mean) radius. That is quite a lot but we are also simulating
the limitations of ``ancient'' astronomy.

To be more concrete, I took 365 values of
the anomaly $\varphi_i=(i-1)/365$, $i=1,\ldots,365$, and for each $i$,
corresponding $50$ values of the radius 
\begin{equation}
    r_{i,l} = \frac{1}{1+0.0167\cos[2\pi\frac{i-1}{365}]} + 
    \epsilon_{i,l}(0,\sigma),\;\; l=1,\ldots,50,
\end{equation}
where the numerical value of eccentricity was used, and $i=1$ coincides with
the minimal radius. Which is not to say, the hypothetical observer knows this
fact. A value of the perihelion anomaly will also have to be found when
fitting.

Next, I calculated, for each value of $\varphi_i$, the mean $r_i$, and its
error
\begin{equation*}
\begin{aligned}
    r_i &= \frac{1}{50}\sum_{l=1}^{50}r_{i,l},\\
    \sigma_i &=
    \sqrt{\frac{1}{50}\frac{1}{49}\sum_{l=1}^{50}(r_{i,l}-r_i)^2}.
\end{aligned}
\end{equation*}

And the respective likelihoods are
\begin{equation}
    L_1 =
    \exp\left[-\frac12\sum_{i=1}^{365}
    \left(\frac{1-r_i}{\sigma_i}\right)^2\right],
\end{equation}
\begin{equation}
    L_2 =
    \exp\left[-\frac12\sum_{i=1}^{365}
    \left(\frac{\frac{1}{1+e\cos(\varphi_i-\varphi_0)}-r_i}
    {\sigma_i}\right)^2\right],
\end{equation}
where a multiplicative constant was omitted for brevity.

Maximising the above, one obtains the values of $L_1$ and $L_2$ required for
formula \eqref{aka} (and, of course $\varphi_0$ and $e$ but these are
unimportant for this experiment). To make sure the result is not just a
coincidence I calculated the mean
$\Delta\rm{AIC} = \rm{AIC}_1 - \rm{AIC}_2$ and its error for $100$ such
observational setups to get
\begin{equation}
    \Delta\rm{AIC} = 8.16 \pm 0.76.
\end{equation}

\section{Conclusions}

Figures 1 and 2 show typical data points (black), together with the fitted elliptic
orbit (blue) and circular orbit (red). AIC gives clear indication in favour of the
ellipse even for such high level of noise. Thus, at least at this point, the
progress of physics would not have been inhibited by model comparing criteria,
and the seemingly more complicated theory would have been chosen. The numbers
and figures speak for themselves, but it is also worth mentioning that if the
errors are reduced only to
$0.1$ the mean $\Delta$AIC increases drastically to the value of $263.3\pm3.3$.
On the other hand, when $\sigma$ is put equal $0.9$, the evidence is 
$\Delta$AIC $=0.73\pm0.53$, which cannot be called
conclusive, but is still positive despite the unrealistic error of 90\% the
radius length.

\begin{figure}
\includegraphics[scale=.75]{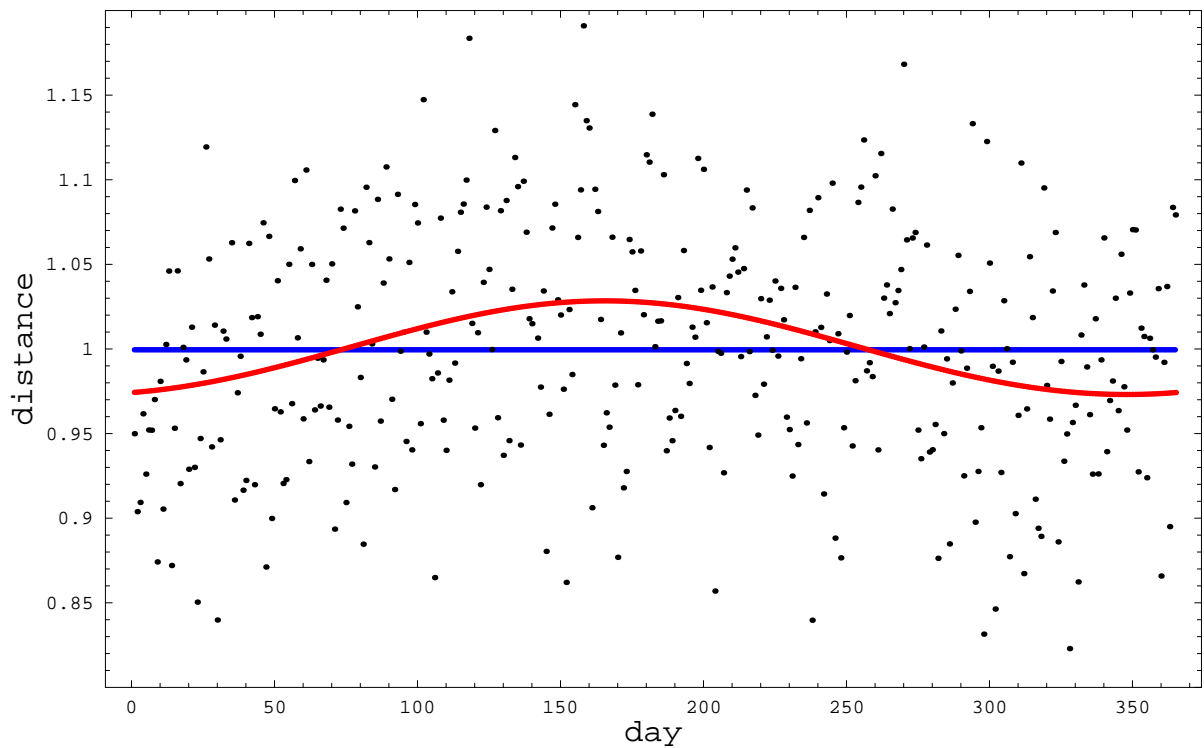}
\caption{Radius plotted against the day number for the simulated
data points (black), the fitted elliptic curve (blue) and the circular orbit
(red).}
\end{figure}
\begin{figure}
\includegraphics[scale=.75]{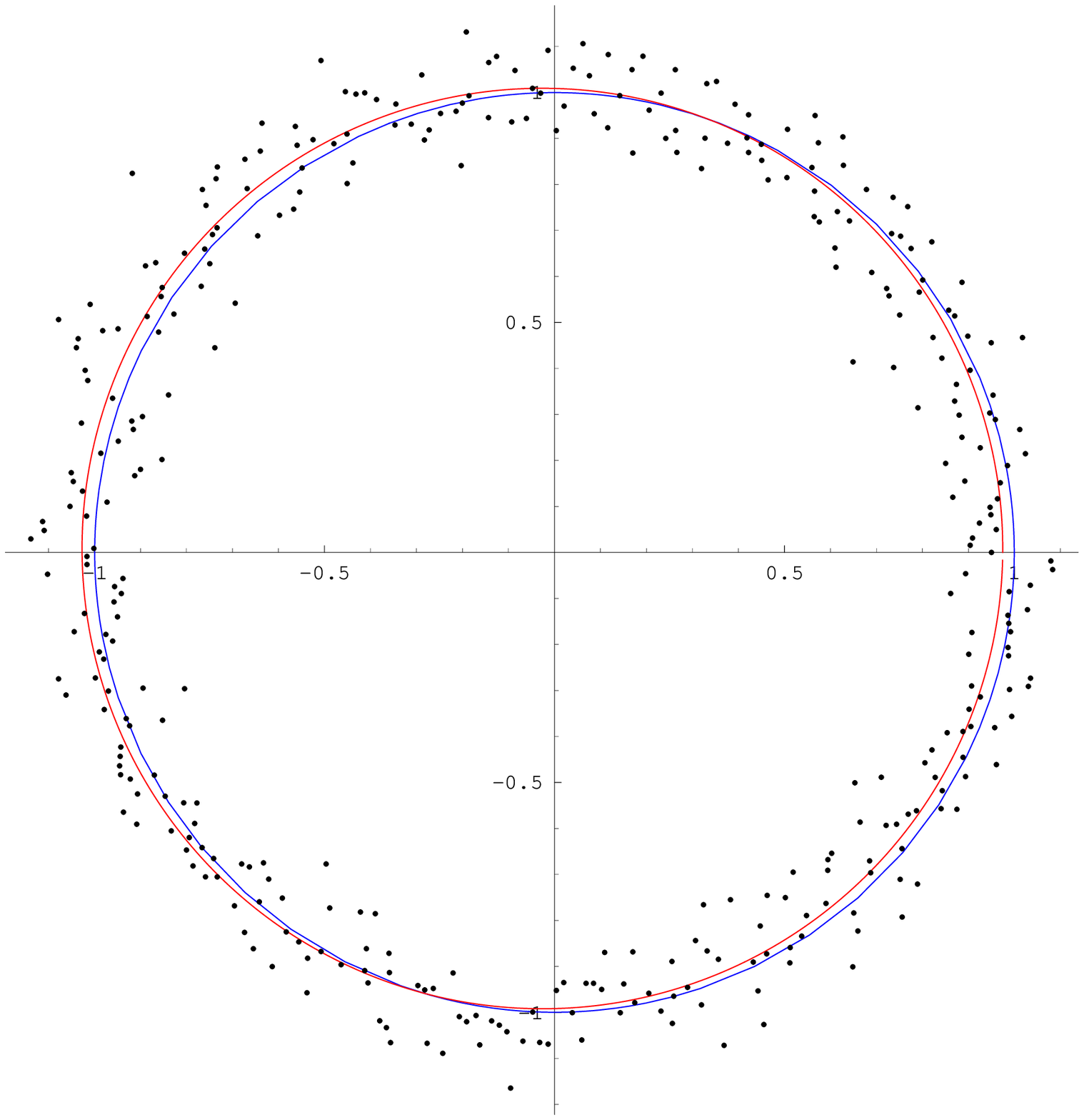}
\caption{Polar plot of the simulated data points (black), fitted elliptic orbit
(blue) and circular orbit of radius $1$ (red).}
\end{figure}

Hopefully, this example will help to understand that model selection
criteria take into account not only the number of parameters but also the
agreement with the data. This is not to say that AIC is
{\it the} criterion of simplicity or elegance of models, but that it still 
gives a reasonable estimate of complexity (parameters) versus applicability
(fitting) of models.

\end{document}